\documentclass[AMA,LATO2COL]{WileyNJD-v2}
\usepackage{moreverb}
\usepackage{bm} %
\newcommand{\beginsupplement}{%
        \setcounter{table}{0}
        \renewcommand{\thetable}{S\arabic{table}}%
        \setcounter{figure}{0}
        \renewcommand{\thefigure}{S\arabic{figure}}%
     } %
\usepackage{fancyhdr} %
\newcommand\BibTeX{{\rmfamily B\kern-.05em \textsc{i\kern-.025em b}\kern-.08em
T\kern-.1667em\lower.7ex\hbox{E}\kern-.125emX}}

\articletype{Research Article}%

\received{<day> <Month>, <year>}
\revised{<day> <Month>, <year>}
\accepted{<day> <Month>, <year>}

\begin{document}

\title{Combining Deep Learning and 3D Contrast Source Inversion in MR-based Electrical Properties Tomography}

\author[1]{Reijer L. Leijsen*}

\author[2,3]{Cornelis A.T. van den Berg}

\author[1]{Andrew G. Webb}

\author[4]{Rob F. Remis}

\author[2,3]{Stefano Mandija}

\authormark{LEIJSEN \textsc{et al}}

\address[1]{\orgdiv{Department of Radiology, C.J. Gorter Center for High Field MRI}, \orgname{Leiden University Medical Center}, \orgaddress{\state{Leiden}, \country{The Netherlands}}}

\address[2]{\orgdiv{Department of Radiotherapy, Division of Imaging \& Oncology}, \orgname{University Medical Center Utrecht}, \orgaddress{\state{Utrecht}, \country{The Netherlands}}}

\address[3]{\orgdiv{Computational Imaging Group for MR diagnostics \& therapy, Center for Image Sciences}, \orgname{Utrecht University}, \orgaddress{\state{Utrecht}, \country{The Netherlands}}}

\address[4]{\orgdiv{Circuits and Systems Group, Faculty of Electrical Engineering, Mathematics and Computer Science}, \orgname{Delft University of Technology}, \orgaddress{\state{Delft}, \country{The Netherlands}}}

\corres{*Reijer Leijsen, C.J. Gorter Center for High Field MRI, Leiden University Medical Center, 2333ZA Leiden, The Netherlands. \email{R.L.Leijsen@lumc.nl}
\vspace{10pt}\\
\textbf{Funding information} \\
European Union's Horizon 2020 Research and Innovation programme, Grant/Award Number: 670629.
\vspace{10pt}\\
\textbf{Word count} \\
3014}


\abstract[Abstract]{Magnetic resonance-electrical properties tomography (MR-EPT) is a technique used to estimate the conductivity and permittivity of tissues from MR measurements of the transmit magnetic field. Different reconstruction methods are available, however all these methods present several limitations which hamper the clinical applicability. Standard Helmholtz based MR-EPT methods are severely affected by noise. Iterative reconstruction methods such as contrast source inversion-EPT (CSI-EPT) are typically time consuming and are dependent on their initialization. Deep learning (DL) based methods require a large amount of training data before sufficient generalization can be achieved. 
Here, we investigate the benefits achievable using a hybrid approach, i.e. using MR-EPT or DL-EPT as initialization guesses for standard 3D CSI-EPT. Using realistic electromagnetic simulations at 3~T and 7~T, the accuracy and precision of hybrid CSI reconstructions are compared to standard 3D CSI-EPT reconstructions.
Our results indicate that a hybrid method consisting of an initial DL-EPT reconstruction followed by a 3D CSI-EPT reconstruction would be beneficial. DL-EPT combined with standard 3D CSI-EPT exploits the power of data driven DL-based EPT reconstructions while the subsequent CSI-EPT facilitates a better generalization by providing data consistency.}


\jnlcitation{\cname{%
\author{Leijsen, R.L.}, 
\author{van den Berg, C.A.T.}, 
\author{Webb, A.G.}, 
\author{Remis, R.F.}, and 
\author{Mandija, S.}} (\cyear{20xx}), 
\ctitle{Combining Deep Learning and 3D Contrast Source Inversion in MR-based Electrical Properties Tomography}, \cjournal{NMR Biomed.}, \cvol{20xx;00:1--x}.}

\maketitle
\thispagestyle{fancy}
\footnotetext{\textbf{Abbreviations:} EM, electromagnetic; EPs, electrical properties; EPT, electrical properties tomography; CSF, cerebrospinal fluid; CSI, contrast source inversion; DL, deep learning; GM, gray matter; MR, magnetic resonance; SNR, signal-to-noise ratio; WM, white matter.}

\section{Introduction}\label{sec1}
Knowledge of \emph{in vivo} tissue electrical properties (EPs; conductivity $\sigma$ and relative permittivity $\varepsilon_\text{r}$) is of high interest for different applications such as improving local specific absorption rate quantification used in e.g. hyperthermia treatment planning or safety assessment in MRI\cite{lagendijk2000hyperthermia,makris2008mri}. Furthermore, due to the relation between conductivity and ionic content, \emph{in vivo} measurements of tissue EPs can in principle provide clinical information about pathological tissues making them a potentially useful biomarker for diagnostic purposes and treatment monitoring \cite{shin2015initial,kim2016correlation}.

There have been several approaches to non-invasively measure \emph{in vivo} tissue electrical properties\cite{katscher2013recent,zhang2014magnetic}. 
In 1991, the possibility to retrieve tissue electrical properties in the radiofrequency range from MR measurements of the circularly polarized transmit magnetic field $(\hat{B}_1^+)$ has been shown\cite{haacke1991extraction}. This technique was referred to as electrical properties tomography (EPT\cite{liu2017electrical,katscher2017electric}).

EPT approaches can be divided into two major categories: direct and inverse methods.
Direct methods based on the Helmholtz equation aim to reconstruct tissue EPs from MR measurement by computing spatial derivatives of the measured $\hat{B}_1^+$ field. However, this operation leads to severe boundary errors and noise amplification in the reconstructed EP maps.

Inverse methods like contrast source inversion-EPT (CSI-EPT\cite{balidemaj2015csi,arduino2017csi,leijsen2018three}) aim at reconstructing EPs by iteratively solving a minimization problem where the EP model is fit to the measured $\hat{B}_1^+$ field. This avoids the need of computing spatial derivatives of measured data, making these methods in principle more noise-robust. 
However, these methods are limited by their computational complexity and the need for EM quantities that are not directly accessible with MRI measurements (such as background fields or transmit phase)\cite{liu2017electrical,katscher2017electric}. Furthermore, CSI-EPT reconstructions suffer from artifacts arising from the low electric field strength at the center of a volume transmit coil and local minima, making CSI-EPT reconstructions dependent upon their initialization. 

Recently, a new approach, called deep learning-EPT (DL-EPT\cite{mandija2019opening}), has been proposed, where the inverse transformation is learned by means of a convolutional neural network. This method relies purely on measurable MR quantities, making it applicable to MR measurements. Preliminary results demonstrated the feasibility of this approach leading to good quality EP maps.
However, the major risk of DL-based EPT reconstructions is that cases not present in the training set will not be accurately reconstructed. Therefore, exhaustive datasets are needed in training, increasing the computational load for DL-based EPT methods.

In this work, a two-step approach is proposed where Helmholtz based reconstructions (MR-EPT) and deep learning reconstructions (DL-EPT) are used as data-driven initializations for 3D CSI-EPT. 
We show that an accurate initialization guess provided by DL-EPT improves CSI-EPT reconstructions, while CSI-EPT has the potential to improve tissue structure of DL-EPT reconstructions.

\section{Methods}
\subsection{EM Simulation Setup} \label{sec:EMsetup}
Electromagnetic (EM) field simulations were performed using commercial finite-difference time-domain EM simulation software XFdtd (Remcom State College, PA, USA). At 3~T a high-pass quadrature birdcage body coil (length of 58~cm, diameter of 70.4~cm) resonant at 128~MHz was simulated, surrounded by a shield (length of 70~cm, diameter of 74.3~cm), while for simulations at 7~T a high-pass quadrature birdcage head coil (length of 19.5~cm, diameter of 30~cm; similar to the dimensions of the Nova Medical birdcage Tx/Rx head coil) resonant at 300~MHz was used, surrounded by a shield (length 22~cm, diameter 36~cm). 
The head of the male human body model (Duke, Virtual Family\cite{christ2009virtual}) used for these simulations was placed at the center of each coil and discretized on a 2x2x2~mm grid. 
For these $\hat{B}_1^+$ field simulations, the object is bounded to the reconstruction domain (128x128x56 voxels), to prevent influences from tissues outside the reconstruction domain on the $\hat{B}_1^+$ field.

\subsection{Reconstruction approaches}\label{sec:reconMethods}
Standard Helmholtz based MR-EPT, DL-EPT and standard 3D CSI-EPT using a homogeneous mask as initialization guess were performed as described below. Furthermore, hybrid reconstructions were performed by providing MR-EPT and DL-EPT reconstructions as initialization guesses to 3D CSI-EPT.

\subsubsection{MR-EPT}
The conventional implementation of MR-EPT is based on the Helmholtz equation, given by
\begin{equation}\label{eq:Helmholtz}
\frac{\nabla^2\hat{B}_1^+}{\hat{B}_1^+}= -\omega^2\mu\varepsilon_0\varepsilon_\text{r} + \text{i}\omega\mu\sigma
\end{equation}
with $\omega$ the Larmor frequency (128~MHz or 300~MHz for 3~T and 7~T, respectively), $\mu$ the permeability of the tissue, which is assumed to be equal to that of free space, and $\varepsilon_0$ the permittivity of free space. If the left-hand side of Equation \ref{eq:Helmholtz} is known, the unknown tissue parameters $\varepsilon_\text{r}$ and $\sigma$ can easily be extracted from this equation. To compute the Laplacian of the simulated $\hat{B}_1^+$ fields in XFdtd, a large 3D finite difference kernel is used (see Mandija et al.\cite{mandija2018error} for its description), since small finite difference kernels are highly sensitive to spatial fluctuations\cite{mandija2018error}.
This approach is hereafter referred to as MR-EPT.

\subsubsection{DL-EPT}
Deep learning-EPT is a data driven approach, where a surrogate model based on accessible MR quantities is learnt. 
Following the procedure indicated in Mandija et al.\cite{mandija2019opening}, deep learning reconstructions for the Duke head model were obtained using a conditional generative adversarial network trained on 1120 unique $\hat{B}_1^+$ fields obtained from realistic EM simulations at 3~T performed in Sim4Life (ZMT, Zurich, Switzerland) with noise superimposed. For these simulations the same body coil simulated in XFdtd (see Section \ref{sec:EMsetup}) and 20 head models (variations of the male and female human head models Duke and Ella from the Virtual Family) were used. 
Moreover, this network was trained using the transceive phase, i.e. the combination of the transmit and receive phase, and not using the transmit phase only, since this latter field can not be measured in an MR experiment\cite{van2014electrical}.
Note that only a deep learning network trained for head models at 3~T is currently available. 
This network provides 2D EP reconstructions, and therefore abrupt changes can appear through slices. 
These reconstructions are referred to as DL-EPT. 

\begin{figure*}
    \centering
    \includegraphics[width=1.8\columnwidth]{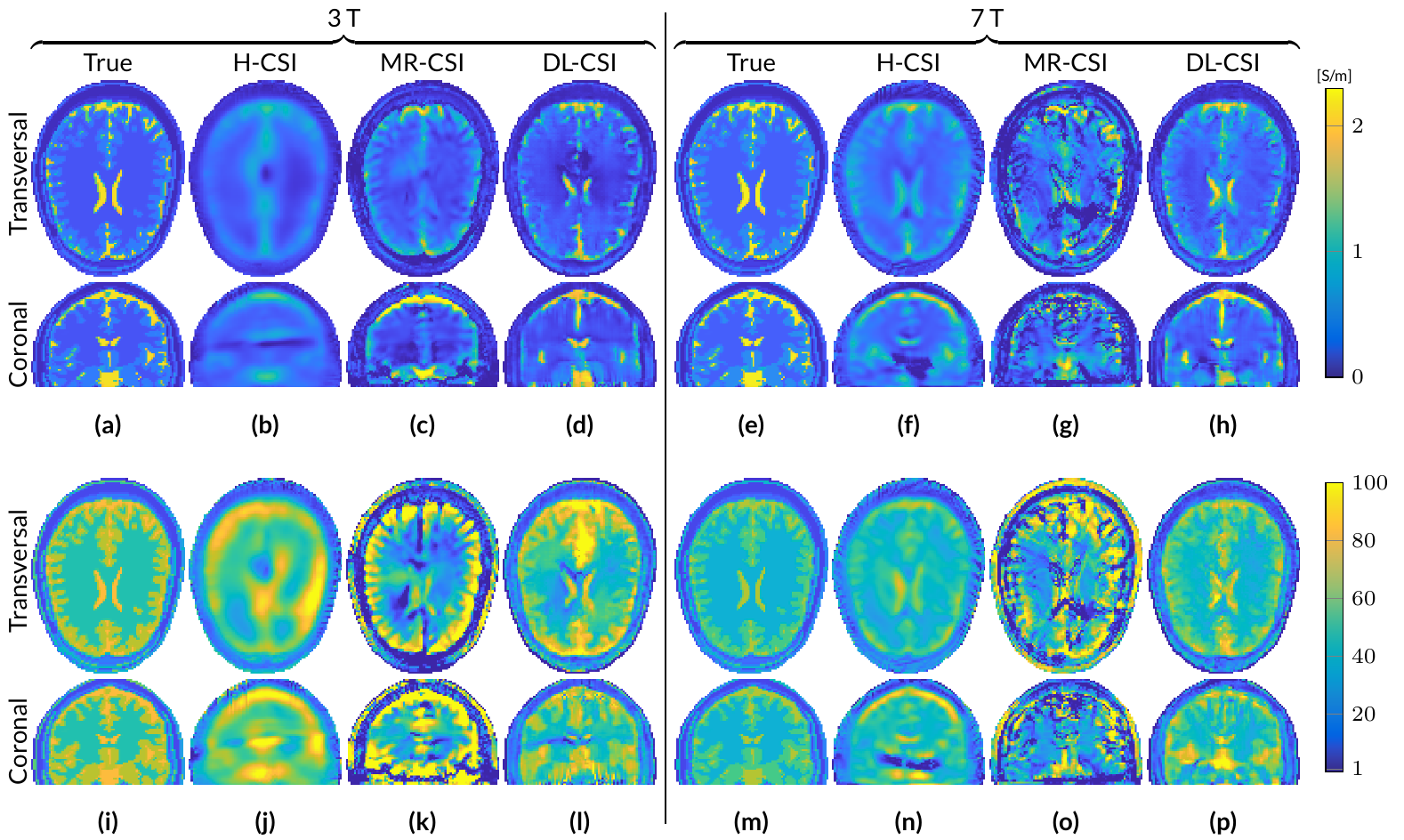}
    \caption{Reconstructed EP maps from different EPT reconstruction approaches for the male head model at 3~T and 7~T based on noiseless $\hat{B}_1^+$ data. Conductivity (a-h) and permittivity (i-p).}
    \label{fig:nonTumor_noiseless}
\end{figure*}
\subsubsection{H-CSI}
Three-dimensional CSI-EPT is an iterative method that minimizes a cost functional based on models of the contrast function $\hat{\chi}$ describing the EPs, and the contrast source $\hat{\bm{w}}=\hat{\chi}\hat{\bm{E}}$, where $\hat{\bm{E}}$ is the electric field strength. The functional that is minimized is given by
\begin{equation}
F(\hat{\bm{w}},\hat{\chi})= \frac{||\hat{\rho}||^2}{||\hat{B}_1^{+;\text{sc}}||^2} + \frac{||\hat{\bm{r}}||^2}{||\hat{\chi}\hat{\bm{E}}^\text{inc}||^2}
\end{equation}
where $\hat{\rho}$ is the mismatch between measured and modeled data, $\hat{\bm{r}}$ describes the discrepancy in satisfying Maxwell's equations, and the superscripts `sc' and `inc' denote the scattered and incident part of the EM fields (see Leijsen et al.\cite{leijsen2018three} for more details). A conjugate gradient update step is used for the contrast function to suppress sensitivity to low electric field regions \cite{leijsen2019developments}. Reconstructions are stopped after 500 iterations or when the functional has reached a tolerance level of $10^{-5}$. These stopping criteria are set to prevent noise overfitting and long reconstruction times. The EPs are extracted from the real and imaginary part of the reconstructed contrast function. 
The $\hat{B}_1^+$ fields simulated in XFdtd were used for standard 3D CSI-EPT reconstructions. For these reconstructions, a homogeneous mask (H) containing the average expected EP values ($\sigma=0.53$ and $\varepsilon_\text{r}=51$ for 3~T reconstructions, $\sigma=0.59$ and $\varepsilon_\text{r}=43$ for 7~T reconstructions) was used as initialization. We refer to these standard CSI reconstructions as H-CSI.

\subsubsection{MR-CSI \& DL-CSI}
As hybrid approaches, we used the performed MR-EPT and DL-EPT reconstructions as initialization masks for 3D CSI-EPT reconstructions. Hereafter we call these two hybrid approaches MR-CSI and DL-CSI.

Note that DL-EPT reconstructions are available only at 3~T. Therefore for DL-CSI reconstructions at 7~T, we used as initialization step the DL reconstruction at 3~T as well.

\subsection{Statistic Evaluation and Constraints}
For all these five methods, EP reconstructions from noiseless data were first performed. Afterwards, the impact of noise on EP reconstructions was investigated. For this purpose, Gaussian noise was added to the real and imaginary parts of the simulated complex $\hat{B}_1^+$ fields in XFdtd leading to a signal-to-noise ratio (SNR) of 100.

Since EP reconstruction methods may lead to voxels with unrealistic EP values, minimum and maximum constraints were applied to the final reconstructions, i.e. bounding the conductivity in the range [0 - 2.6]~S/m and the permittivity in the range [1 - 100], where the maximum values are approximately 20\% higher than the maximum EP values present in the ground truth dielectric models.

For all these reconstructions, mean and standard deviation values were computed in the white matter (WM), gray matter (GM), and cerebrospinal fluid (CSF) regions as a proxy of accuracy and precision of these reconstruction methods.
Moreover, to evaluate the overall reconstruction accuracy among the investigated EP reconstruction methods, the global relative residual error was computed as
\begin{equation}
\text{RRE}=\frac{||\hat{x}-\tilde{x}||}{||\hat{x}||},
\end{equation}
where $\hat{x}$ depicts the true conductivity or relative permittivity and $\tilde{x}$ the reconstructed one, and the norm is the Euclidean norm defined over the complete domain of interest.

\begin{figure*}
    \centering
    \includegraphics[width=1.8\columnwidth]{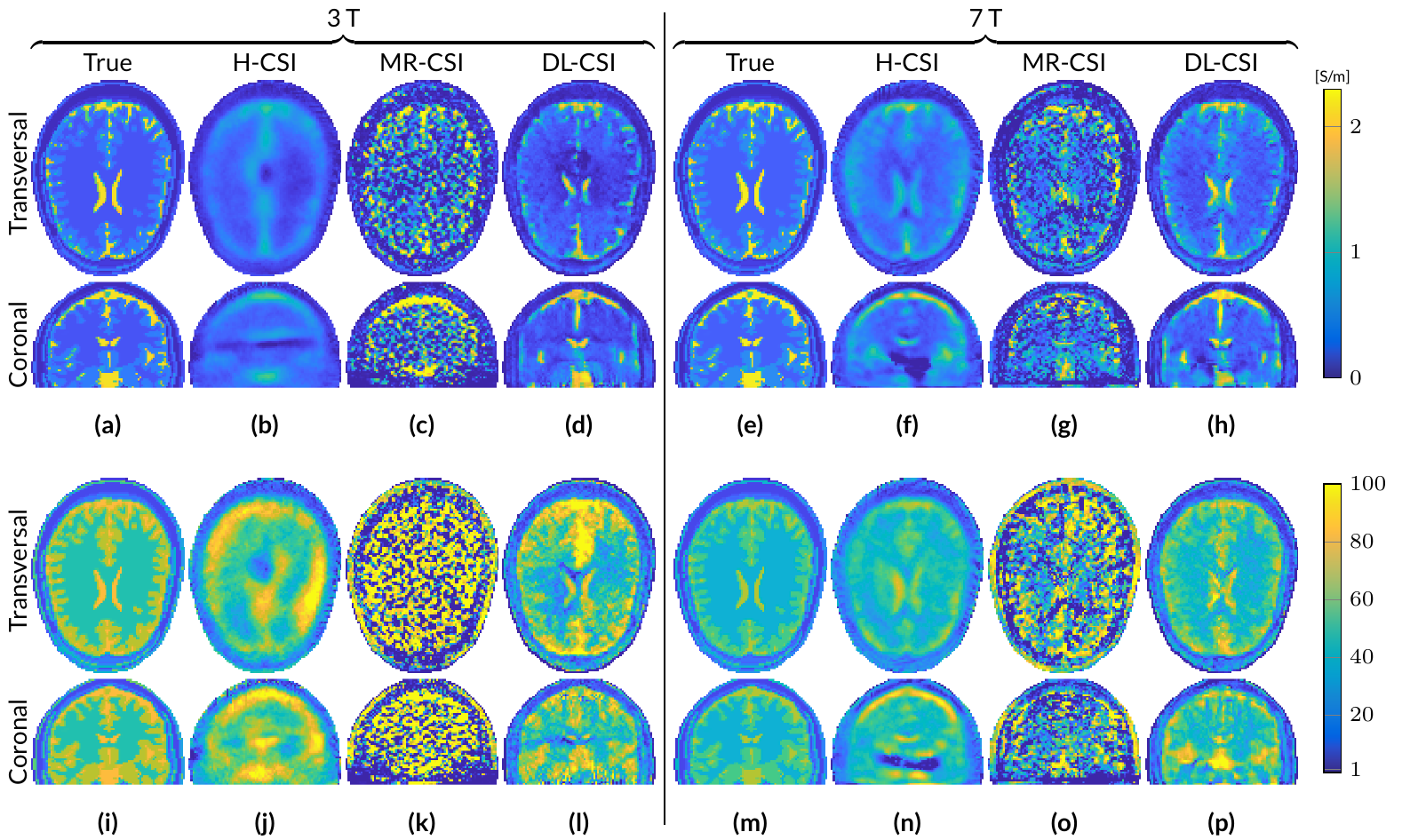}
    \caption{Reconstructed EP maps from different EPT reconstruction approaches for the male head model at 3~T and 7~T based on $\hat{B}_1^+$ data with an SNR of 100. Conductivity (a-h) and permittivity (i-p).}
    \label{fig:nonTumor}
\end{figure*}
\section{Results}
Figure 1 shows EP reconstructions at 3~T and 7~T for noiseless simulated $\hat{B}_1^+$ data using the standard 3D CSI-EPT (H-CSI) and the two hybrid CSI-EPT approaches (MR-CSI and DL-CSI). All the results are obtained after 500 CSI iterations (taking around 60 minutes on a standard computer for this reconstruction domain size), at which point the mismatch functional has decreased to a value of about $5 \times 10^{-5}$.

At 3~T, H-CSI produces very poor EP maps. The conductivity map shows a smooth reconstruction with underestimations of the high conductivity values. The permittivity map shows less of the underlying tissue structure and the white matter region contains clear overestimations. Furthermore, distorted reconstructions are observed in the center of the object in both EP maps, corresponding to the region with a low electric field strength. 
For MR-CSI conductivity reconstructions, an improvement is observed especially at the periphery of the head, i.e. away from the low $|\hat{\bm{E}}|$-field region. However, in the low $|\hat{\bm{E}}|$-field region (ventricles), conductivity reconstructions are still erroneous. MR-CSI permittivity reconstructions shows severe distortions throughout the brain, reflecting the severe boundary errors of standard MR-EPT reconstructions (see Figure S1). 
The DL-CSI approach shows better tissue reconstructions of the ventricles compared to H-CSI and MR-CSI. They are clearly visible in both the conductivity and permittivity map. However, even though the reconstructed DL-CSI EP values are close to the ground truth values, small errors arising from the DL-EPT reconstructions used as initialization step (see Figure S1) are visible at the periphery of the head. Note that the DL-CSI reconstructions presented in this figure assumes noiseless data, while DL-EPT reconstructions used as initialization for DL-CSI are available only for noisy data, since the available neural network was trained only on the noisy $\hat{B}_1^+$ data to better resemble realistic scenarios from MR-measurements.

At 7~T, similar results are observed for H-CSI as at 3~T: a smoothed version for the reconstructed conductivity, and overestimations in the homogeneous WM region for the reconstructed permittivity. Also the low $|\hat{\bm{E}}|$-field region is clearly visible, which at 7~T is located further down compared to the 3~T case (compare the coronal slices of Figure 1b,j with 1f,n). 
MR-CSI at 7~T shows boundary artifacts, which are the result of the intrinsic errors of MR-EPT at tissue boundaries. However, compared to its initialization (Figure S1), improvements are observed in the conductivity maps. 
DL-CSI reconstructions at 7~T (using as initialization DL-EPT reconstructions at 3~T) show higher structure fidelity compared to DL-EPT reconstructions, especially around the ventricles.

\begin{figure*}
    \centering
    \includegraphics[width=2.0\columnwidth]{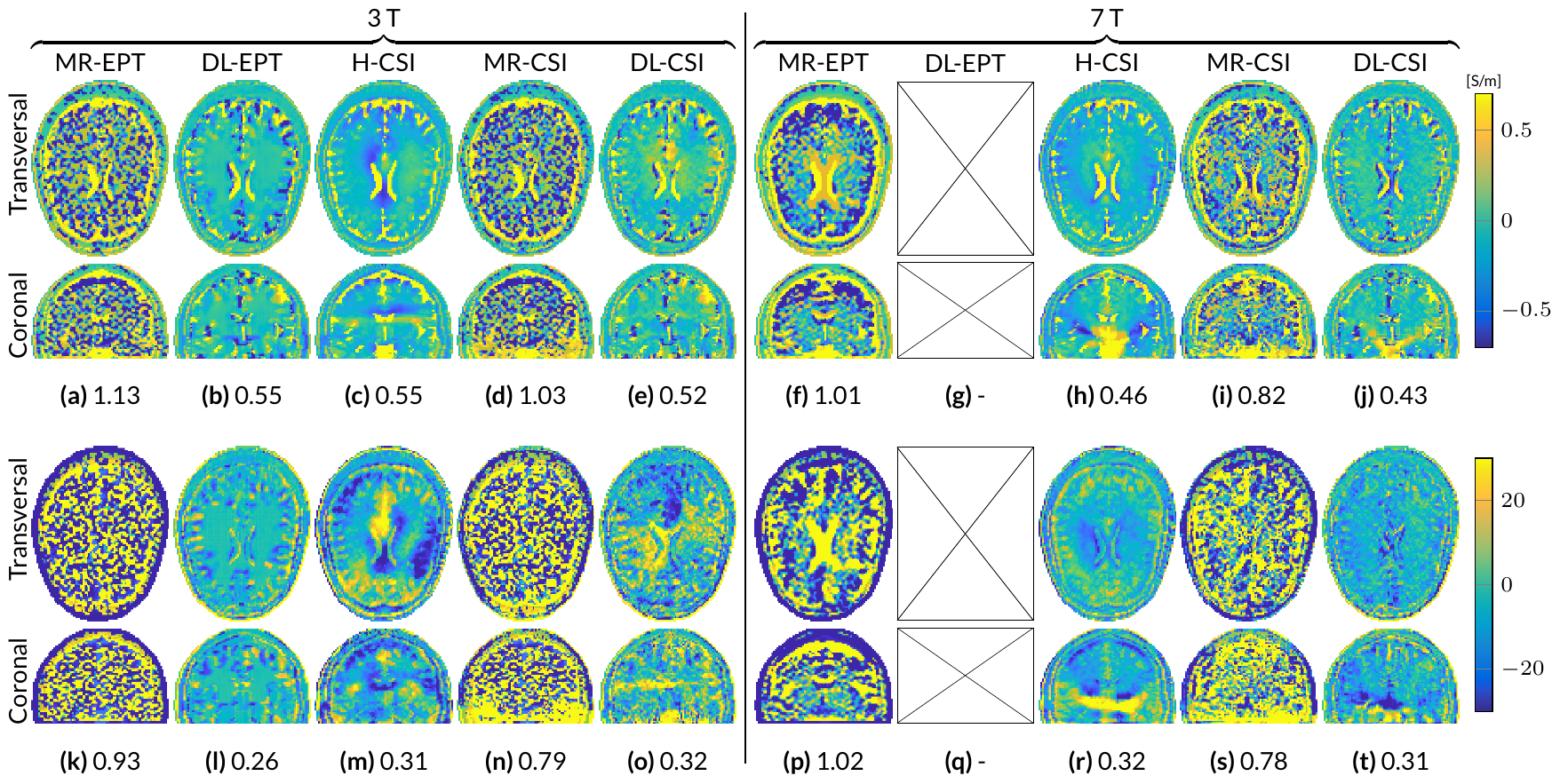}
    \caption{Absolute error maps (ground truth - reconstruction) of the reconstructions from the different EPT approaches, for the Duke head model at 3~T and 7~T. The values in the subcaptions denote the RRE of the whole volume. Conductivity (a-j) and permittivity (k-t).}
    \label{fig:nonTumor_error}
\end{figure*}
Figure 2 shows reconstruction results when Gaussian noise is present in the $\hat{B}_1^+$ data $(\text{SNR}=100)$. 
The introduction of noise results in non-significant differences for H-CSI reconstructions both at 3~T and 7~T. 
This is in contrast to MR-CSI, whose initialization maps, i.e. EP maps obtained from standard Helmholtz MR-EPT reconstructions, are extremely sensitive to noise (see Figure S1), thus leading to noise corrupted MR-CSI reconstructions, both at 3~T and 7~T.
DL-CSI reconstructions are minimally affected by noise, which leads to slightly higher standard deviations in permittivity reconstructions for DL-CSI compared to DL-EPT (See Table \ref{tab:nonTumor}).

In the supplementary material Table \ref{tab:nonTumor}, the mean and standard deviation values are reported for EP reconstructions in the WM, GM and CSF from noiseless and noisy data for all the aforementioned methods (MR-EPT, DL-EPT, H-CSI, MR-CSI and DL-CSI) as a proxy of accuracy and precision.

Figure 3 gives a qualitative impression of the reconstruction errors of the five different reconstruction methods, by showing the absolute error maps for EP reconstructions from noisy $\hat{B}_1^+$ data.
For a direct quantitative comparison, the computed RRE in the whole domain is also reported in the figure for each reconstruction method.

MR-EPT reconstructions show severe errors due to noise amplification in the reconstructed EP maps at 3~T. These errors are reduced for 7~T MR-EPT reconstructions. However, the quality of the reconstructed EP maps remains still poor.
DL-EPT reconstructions, available only at 3~T, show good accuracy in homogeneous regions. However, reconstruction errors are present at tissue boundaries, e.g. around the ventricles.
H-CSI shows substantial errors arising from the low $|\hat{\bm{E}}|$-field region. These errors appear as artificial bands/shadow artifacts, which are intrinsically caused by the homogeneous initialization.
MR-CSI is strongly affected by the reconstruction errors present in MR-EPT, which is used as initialization guess. Although the RRE is lower compared to MR-EPT reconstructions, the quality of MR-CSI is still poor.
DL-CSI reconstructions show good quality EP maps. The combined conductivity and permittivity RRE values of DL-CSI are lower compared to H-CSI both at 3~T and 7~T, showing an advantage in using DL as initialization guess. Artifacts such as artificial bands present in H-CSI reconstructions are highly reduced in DL-CSI. Improvements are also observed with respect to DL-EPT (see Figure S1), especially for conductivity reconstructions around the ventricles.

\section{Discussion \& Conclusion}
In this manuscript, we investigated the possible benefits for EP reconstructions achievable by combining standard MR-EPT, DL-EPT, and 3D CSI-EPT (H-CSI) into a hybrid reconstruction approach, i.e. by providing MR-EPT or DL-EPT reconstructions as an initialization guess for CSI-EPT. By doing so, CSI provides data consistency for MR-EPT or DL-EPT reconstructions, i.e. the data needs to satisfy Maxwell's equations, while MR-EPT and DL-EPT reconstructions provide in principle a better initialization guess for CSI-EPT compared to the standard approach, which uses a simple homogeneous mask.

Reconstructions obtained with CSI-EPT depend on the map provided as initialization guess. If a homogeneous mask is provided (H-CSI), sharp tissue boundaries are in principle reconstructed for a noiseless situation only after a larger number of iterations (e.g. about 10,000\cite{leijsen2018three}). However, for realistic cases including noise, a large number of iterations leads to noise overfitting. To limit this, less iterations are performed, resulting in smoother EP maps. Moreover, H-CSI can converge to sub-optimal results, and regions of very low values in the reconstructed EP maps can occur, even in noiseless cases. It is therefore critical to provide CSI-EPT with a good initialization guess. 

If available methods like MR-EPT are used as the initial guess, improvements can be observed in noiseless cases. 
However, for realistic scenarios $(\text{SNR}=100)$, MR-EPT reconstructions are severely affected by noise. As shown in this work, this also corrupts MR-CSI reconstructions.

DL-EPT is more noise robust than MR-EPT as the adopted network was trained on noisy data. 
However, a major problem for DL-EPT is the need for a large training data set. 
By providing DL-EPT reconstructions as input for CSI, more accurate conductivity reconstructions are obtained at 3~T and 7~T compared to the other methods presented in this work. DL-CSI is able to provide data consistency, which, in principle, can allow higher fidelity in the reconstructed tissue structures. 
For permittivity reconstructions we did not observe a substantial improvement in the reconstruction accuracy for DL-CSI compared to DL-EPT. Still, DL-CSI reconstructions at 7~T show an improvement compared to DL-CSI reconstructions at 3~T, indicating the benefit of high field strengths MRI for permittivity reconstructions.

A limitation in this work is that DL-EPT reconstructions were not available at 7~T, since the network was only trained on simulated noisy data at 3~T. We believe that by using in the future 7~T DL-EPT reconstructions, the DL-CSI reconstructions can further improve due to the higher imprinting of the EPs in the measured fields leading to higher field curvatures with increasing field strengths, similarly to the improvements already shown by H-CSI at 7~T compared to 3~T.

In conclusion, the combination of noise-robust DL-EPT reconstructions and 3D CSI-EPT reconstructions allows in principle better generalization since CSI-EPT introduces data consistency for the subject at hand. 
This might reduce the need of an exhaustive training dataset for DL-EPT. 
Meanwhile, using DL-EPT as initialization for CSI-EPT improves the quality and accuracy of standard 3D CSI-EPT reconstructions.

\section*{Acknowledgements}
This research was partially funded by European Research Council Advanced NOMA MRI under grant number 670629. (R.~L., A.~W.) We thank E.~F.~Meliad\`o, P.~R.~S. Stijnman and N.~R.~F.~Huttinga for their support to this work.

\bibliography{wileyNJD-AMA}%


\clearpage
\section*{Supplementary Material}
\beginsupplement
\vfill 
\begin{minipage}{2\columnwidth}
\captionsetup{type=figure}
    \centering
    \includegraphics[width=0.9\columnwidth]{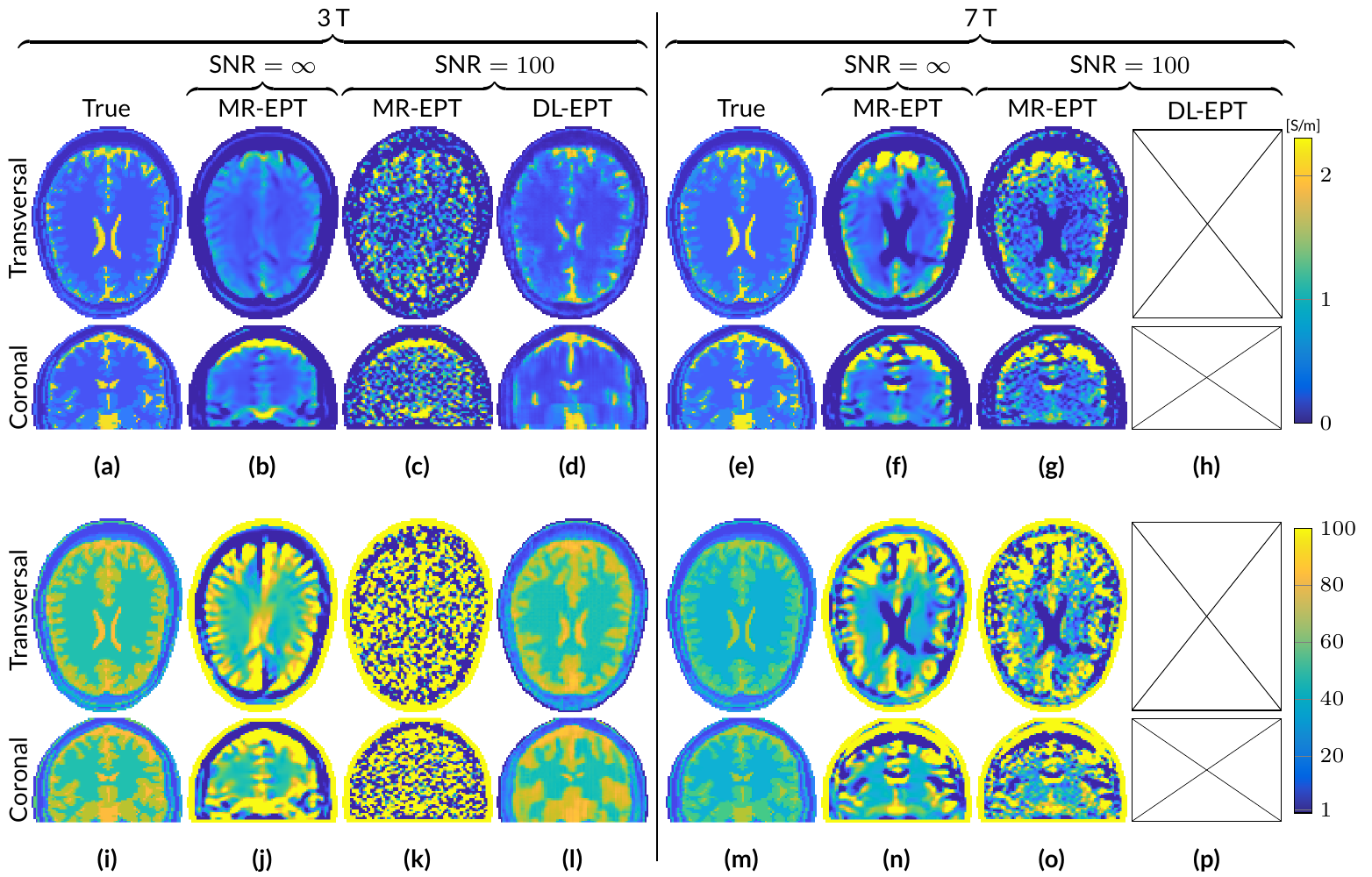}
    \caption{MR-EPT and DL-EPT reconstructions from noiseless and noisy 3~T and 7~T $\hat{B}_1^+$ data of the Duke head model. These reconstruction are used as initialization for MR-CSI and DL-CSI. Note that 7~T DL-EPT reconstructions are not available. Conductivity (a-h) and permittivity (i-p).}
    \label{fig:EPTappendix}
\end{minipage}
\vfill

\vfill
\begin{table*}[h]
\caption{The mean and standard deviation of the reconstructed EPs for the segmented regions white matter (WM), gray matter (GM) and cerebrospinal fluid (CSF) for the Duke head model without and with noise at 3~T and 7~T. The difference in averages between the noiseless and noisy case in MR-EPT (and thus MR-CSI) is due to the applied minimum and maximum constraint.}
\label{tab:nonTumor}
\center
    \begin{tabular}{llcccccc}
        \toprule
                                        &         & \multicolumn{6}{c}{Conductivity}\\
                                        \cmidrule(r){3-8}
                                        &         & \multicolumn{3}{c}{3~T} & \multicolumn{3}{c}{7~T}\\                
                                        \cmidrule(r){3-5} \cmidrule(r){6-8}
                                        &         & WM                & GM                & CSF               & WM                & GM                & CSF               \\
        \midrule
                                        & True    & 0.34              & 0.59              & 2.14              & 0.41              & 0.69              & 2.22              \\
        \midrule
        \multirow{4}{*}{SNR $\infty$}   & MR-EPT  & 0.49$\pm$0.27     & 0.70$\pm$0.51     & 0.72$\pm$0.79     & 0.65$\pm$0.54     & 0.86$\pm$0.70     & 0.79$\pm$0.86     \\
                                        & H-CSI   & 0.44$\pm$0.16     & 0.57$\pm$0.16     & 0.92$\pm$0.30     & 0.50$\pm$0.16     & 0.72$\pm$0.24     & 1.39$\pm$0.46     \\
                                        & MR-CSI  & 0.41$\pm$0.29     & 0.69$\pm$0.53     & 0.89$\pm$0.78     & 0.49$\pm$0.40     & 0.66$\pm$0.55     & 0.92$\pm$0.70     \\
                                        & DL-CSI  & 0.41$\pm$0.22     & 0.64$\pm$0.36     & 1.46$\pm$0.55     & 0.46$\pm$0.18     & 0.75$\pm$0.30     & 1.67$\pm$0.51     \\
        \midrule
        \multirow{5}{*}{SNR 100}        & MR-EPT  & 0.66$\pm$0.69     & 0.82$\pm$0.82     & 0.78$\pm$0.90     & 0.67$\pm$0.64     & 0.87$\pm$0.75     & 0.81$\pm$0.89     \\
                                        & DL-EPT  & 0.43$\pm$0.25     & 0.70$\pm$0.41     & 1.54$\pm$0.61     & -                 & -                 & -                 \\        
                                        & H-CSI   & 0.44$\pm$0.16     & 0.57$\pm$0.16     & 0.90$\pm$0.30     & 0.51$\pm$0.17     & 0.72$\pm$0.25     & 1.32$\pm$0.46     \\
                                        & MR-CSI  & 0.64$\pm$0.65     & 0.75$\pm$0.77     & 0.87$\pm$0.89     & 0.51$\pm$0.50     & 0.65$\pm$0.60     & 0.94$\pm$0.73     \\
                                        & DL-CSI  & 0.41$\pm$0.23     & 0.64$\pm$0.36     & 1.46$\pm$0.56     & 0.46$\pm$0.19     & 0.75$\pm$0.30     & 1.68$\pm$0.51     \\
        \bottomrule
        \toprule
                                        &         & \multicolumn{6}{c}{Permittivity}\\
                                        \cmidrule(r){3-8}
                                        &         & \multicolumn{3}{c}{3~T} & \multicolumn{3}{c}{7~T}\\                
                                        \cmidrule(r){3-5} \cmidrule(r){6-8}
                                        &         & WM                & GM                & CSF               & WM                & GM                & CSF               \\
        \midrule
                                        & True    & 52.53             & 73.52             & 84.04             & 43.78             & 60.02             & 72.73             \\
        \midrule
        \multirow{4}{*}{SNR $\infty$}   & MR-EPT  & 62.80$\pm$20.40   & 70.94$\pm$32.31   & 56.48$\pm$43.06   & 47.45$\pm$25.16   & 53.91$\pm$33.01   & 62.34$\pm$40.05   \\
                                        & H-CSI   & 60.98$\pm$11.82   & 70.16$\pm$12.81   & 81.43$\pm$13.84   & 46.37$\pm$9.86    & 53.44$\pm$12.97   & 68.32$\pm$18.19   \\
                                        & MR-CSI  & 53.58$\pm$26.93   & 67.02$\pm$34.80   & 60.81$\pm$40.55   & 36.66$\pm$24.61   & 39.66$\pm$30.70   & 51.61$\pm$36.02   \\
                                        & DL-CSI  & 54.83$\pm$11.68   & 68.37$\pm$12.60   & 79.39$\pm$15.00   & 49.34$\pm$8.32    & 61.75$\pm$11.15   & 69.85$\pm$14.08   \\
        \midrule
        \multirow{2}{*}{SNR 100}        & MR-EPT  & 55.64$\pm$43.86   & 60.71$\pm$43.92   & 54.40$\pm$45.69   & 47.60$\pm$30.55   & 53.90$\pm$35.27   & 62.00$\pm$40.71   \\
                                        & DL-EPT  & 58.06$\pm$9.06    & 71.10$\pm$8.68    & 80.40$\pm$6.53    & -                 & -                 & -                 \\        
                                        & H-CSI   & 60.28$\pm$12.10   & 69.80$\pm$13.25   & 80.60$\pm$14.30   & 47.02$\pm$10.93   & 52.42$\pm$13.88   & 67.96$\pm$19.04   \\
                                        & MR-CSI  & 47.85$\pm$41.79   & 46.01$\pm$42.58   & 50.65$\pm$43.44   & 35.97$\pm$29.49   & 37.84$\pm$32.74   & 50.72$\pm$37.35   \\
                                        & DL-CSI  & 54.72$\pm$13.70   & 68.19$\pm$14.53   & 79.00$\pm$16.39   & 49.27$\pm$9.01    & 61.75$\pm$11.67   & 69.81$\pm$14.45   \\
        \bottomrule
    \end{tabular}
\end{table*}
\vfill

\end{document}